\documentclass[prl,twocolumn,superscriptaddress]{revtex4}
\usepackage[normalem]{ulem}
\usepackage{amsmath,amssymb,times,graphicx}
\usepackage[ps2pdf,bookmarks=true,colorlinks,linkcolor=red,urlcolor=blue,citecolor=blue]{hyperref}
\usepackage[usenames]{color}

\newcommand{\be}{\begin{equation}}
\newcommand{\beq}{\begin{eqnarray}}
\newcommand{\eeq}{\end{eqnarray}}
\newcommand{\ms}{m_s}

\def \be{\begin{equation}}
\def \ee{\end{equation}}
\def \ba{\begin{array}}
\def \ea{\end{array}}
\def \bea{\begin{eqnarray}}
\def \eea{\end{eqnarray}}

\def \half{\frac{1}{2}}

\def \D{{\Delta}}

\def \w{{\omega}}

\def \ket#1{{\,|\,#1\,\rangle\,}}

\def \parend{.}
\def \figwa{0.36}
\def \figw{0.37}
\def \figpreamble{}
\newcommand{\fribourg}{Department of Physics, University of Fribourg, CH-1700 Fribourg, Switzerland}
\newcommand{\harvard}{Department of Physics, Harvard University, Cambridge, MA 02138}

\begin{document}

\title{Relaxation of antiferromagnetic order in spin-1/2 chains following a quantum quench}

\author{Peter Barmettler}
\affiliation{\fribourg}
\author{Matthias Punk}
\affiliation{Department of Physics, Technical University Munich, D-85748 Garching, Germany}
\author{Vladimir Gritsev}
\affiliation{\fribourg}
\affiliation{\harvard}
\author{Eugene Demler}
\affiliation{\harvard}
\author{Ehud Altman}
\affiliation{Department of Condensed Matter Physics, Weizmann Institute of Science, Rehovot, 76100, Israel}

\date{January 30, 2009}

\begin{abstract}
We study the unitary time evolution of antiferromagnetic order in anisotropic Heisenberg chains that are initially prepared in a pure quantum state far from equilibrium. Our analysis indicates that the antiferromagnetic order imprinted in the initial state vanishes exponentially. Depending on the anisotropy parameter, oscillatory or non-oscillatory relaxation dynamics is observed. Furthermore, the corresponding relaxation time exhibits a \emph{minimum} at the critical point, in contrast to the usual notion of critical slowing down, from which a maximum is expected.
\end{abstract}
\maketitle

\paragraph{Introduction\parend}
Experiments with ultracold atoms offer a highly controlled environment
for investigating open questions of quantum magnetism. In particular, coherent spin dynamics in a lattice of double wells has been observed in recent experiments, which have demonstrated remarkable precision in tuning magnetic exchange interactions~\cite{trotzky-2008}. The ability to observe quantum dynamics over long time intervals allows one to study strongly correlated states from a new perspective. The idea is to prepare the system in a simple quantum state which, in general, is not an eigenstate of the Hamiltonian, and investigate the dynamics that follows. In the two-spin system, studied in~\cite{trotzky-2008}, the dynamics is completely tractable and describes simple oscillations between a singlet and a triplet states.

In the present paper we investigate how the nature of the dynamics changes in the case of a macroscopic number of spins interacting via nearest neighbor magnetic exchange. Are there new effects, and in particular new time scales, dynamically generated by the complex many-body evolution? Our starting point for investigating this question is the spin-$\half$ anisotropic Heisenberg (or XXZ) model on a one-dimensional lattice
\be
 H_\text{XXZ} =  J\sum_j \left\{ S_j^x S_{j+1}^x + S_j^y S_{j+1}^y + \Delta  S_j^z S_{j+1}^z \right\}\,.
\label{XXZhamiltonian}
\ee
This model provides a good effective description of two-component Bose or Fermi systems deep in the Mott insulating phase. The interaction parameters are dynamically tunable ~\cite{mapping}, realizing ferro- ($J<0$) or antiferromagnetic ($J>0$) couplings over large ranges of the anisotropy parameter $\Delta\geq 0$. We take the initial state to be a perfect antiferromagnetic (N\'eel) state $|\psi_0\rangle=|\uparrow \downarrow \uparrow \dots \downarrow \uparrow \downarrow \rangle$.  Such a state has been achieved with high fidelity by Trotzky {\it et al.} \cite{trotzky-2008} using decoupled double wells. Note that $\ket{\psi_0}$ is the ground state of the Hamiltonian with $\D=\infty$. We study the subsequent time evolution of the staggered magnetic moment
$
\ms(t)= \frac{1}{N} \sum_j (-1)^j \langle \psi_0|S_j^z(t)|\psi_0\rangle\,
$
under the influence of the Hamiltonian (\ref{XXZhamiltonian}) at different values of anisotropy $\D$ using a numerical matrix-product method \cite{vidal-2007}. The dynamics is independent of the sign of $J$ and the results are valid for both ferro- and antiferromagnetic couplings. To substantiate our findings, we consider another, closely related model, given by the XZ-Hamiltonian [see Eq. (\ref{XZhamiltonian})], which allows exact calculation of the dynamics and displays similar behavior.

Theoretical interest in this class of problems, known as quantum quenches \cite{barouch-1,low-energy,boson,cramer-2008,spin,kollar}, has been invigorated by advances in experiments with ultracold atoms \cite{experiments}. In particular, macroscopic order parameter oscillations have been predicted to occur following a quantum quench in a variety of such systems \cite{AltmanAuerbach2002, order-parameter, sengupta-2004,hastings-2008}. We shall see that such oscillations are also found in the XXZ-chain with easy-plane anisotropy ($\D<1$), and that they are essentially the same as the singlet-triplet oscillations observed in the two-spin system~\cite{trotzky-2008}. Accordingly, the oscillation frequency is directly related to the magnetic exchange interaction $J$. More importantly, for non-zero $\Delta$ we find a fundamentally new mode of many-body dynamics which always leads to {\em exponential} decay of the staggered moment regardless of whether the short-time dynamics is oscillatory or not. In contrast with the oscillation frequency, the relaxation time is an emergent scale generated by the highly correlated dynamics and hence cannot be simply related to the microscopic parameters.
We find a diverging relaxation time in the two limits $\D\to 0$ and $\D\to\infty$. Of particular interest is the relaxation time at the isotropic point $\D=1$, which for the ground state properties marks a quantum phase transition from a gapless "Luttinger liquid" phase ($\D<1$) to a gapped, Ising-ordered antiferromagnetic phase ($\D>1$). Interestingly, the relaxation time is \emph{minimal} in the vicinity of the critical point, where its value is simply determined by the magnetic exchange interaction $\tau\sim 1/J$. This accelerated relaxation stands in remarkable contrast to the notion of {\it critical slowing down}, valid for a small perturbation of the order parameter from equilibrium. In fact, if the prepared initial state is close to the equilibrium state, then the relaxation time of the order parameter is expected to diverge as the system approaches the critical point \cite{critical-slowing-down}. We find an opposite trend in the dynamics of the prepared N\'eel state. In the long-time limit, our results suggest that local magnetic order vanishes for all values $\Delta<\infty$.

The solution of the quench dynamics in the XXZ-chain involves in principle all energy scales of the Hamiltonian and approximative methods become essentially inaccurate in many cases. The mean field approximation for example leads to contradictions with our results -- an algebraic decay for $\D\leq1$ and a non-vanishing asymptotic value of the staggered moment for $\D>1$ \cite{hastings-2008}. Renormalization group based approaches \cite{low-energy} are restricted to low-lying modes, which is not sufficient in the present case. The exact numerical results presented in this study go further than the predictions of low-energy theories \cite{low-energy}.

\begin{figure}[t]
\includegraphics[width=\figwa\textwidth,angle=0]{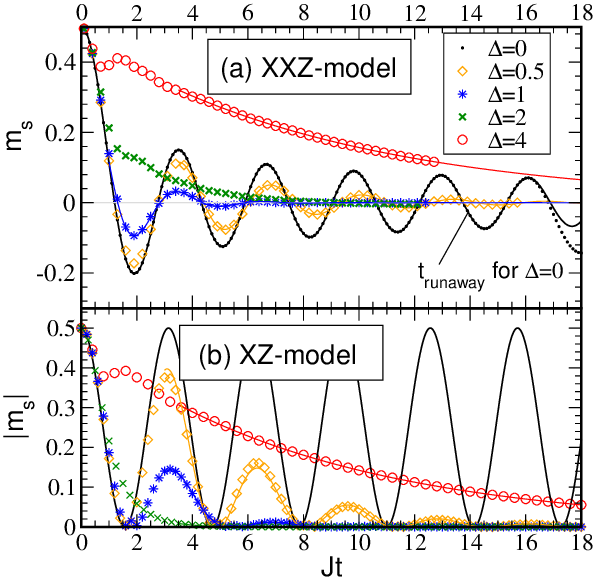}
\caption{\label{fig:ms_all}\figpreamble Dynamics of the staggered magnetization $\ms(t)$ in the XXZ-chain (a) and the XZ-chain (b). Symbols  correspond to numerical results, lines represent analytical results or fits by corresponding laws (see text). For $\Delta=0$ the typical behavior of the error is illustrated by comparing the numerical iTEBD result with 2400 retained states to the exact curve: the absolute deviation from the exact curve is less than $10^{-6}$ for $t<t_{runaway}$. For $\Delta\neq0$ data beyond $t_{runaway}$ is omitted.
}
\end{figure}

Before delving into a more detailed study it is instructive to consider the so-called XX-limit ($\Delta=0$) of the Heisenberg chain (\ref{XXZhamiltonian}), which can be mapped onto the problem of free fermions. In this case one easily obtains an analytic expression for the time evolution of the staggered magnetization: $m_s(t)=J_0(2Jt)/2$ (Fig. \ref{fig:ms_all}).
Here $J_0$ denotes the zeroth-order Bessel function of the first kind. Thus, after a short transient time $t \sim J^{-1}$, the staggered magnetization displays algebraically decaying oscillations originating from the finite bandwidth of the free-fermionic model, 
\beq
\ms(t)\sim\frac{1}{\sqrt{4\pi t}}\cos(2Jt-\frac{\pi}{4})\,.
\eeq
In general, we are interested in generic behavior of the relaxation dynamics on large time scales. We adopt a definition of relaxation which does not rely on time-averaged equilibration of the observable, but instead requires {\it exact} convergence to the asymptotic value, as defined in Ref. \cite{cramer-2008a}. From this point of view, the oscillations in the XX-limit are characterized by an infinite relaxation time.

\paragraph{XXZ-model\parend}

\begin{figure}[t]
\includegraphics[width=\figw\textwidth,angle=0]{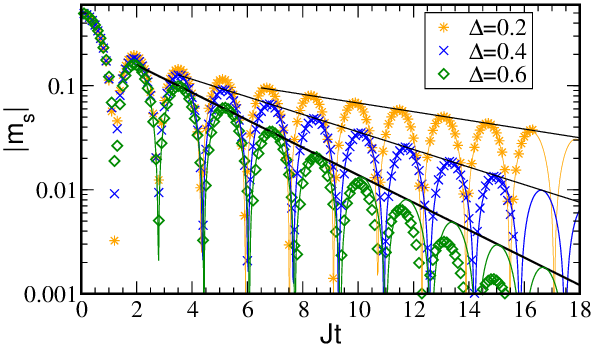}
\caption{\label{fig:ms_weak}\figpreamble Absolute value of the staggered magnetization in the XXZ-model. Symbols represent numerical results, solid curves correspond to fits by the exponential law (\ref{eq:damping}), straight lines point out the exponential decay.
}
\end{figure}

\begin{figure}[t]
\includegraphics[width=\figw\textwidth,angle=0]{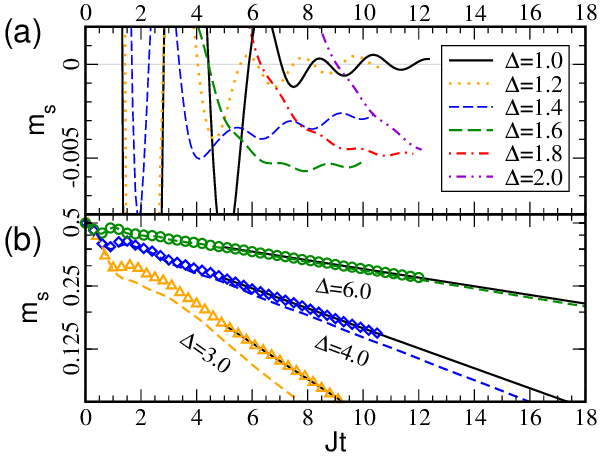}
\caption{\label{fig:ms_strong}\figpreamble (a) Focus on the XXZ-chain close to the critical point $\Delta=1$. (b) Comparison of the XXZ-chain (symbols) and the XZ-chain (dashed lines) for strong anisotropies, solid lines correspond to an exponential fit.
The dynamics of the staggered magnetization of the XXZ- and XZ-chains converge towards each other in the large-$\Delta$ limit.
}
\end{figure}

In the general case of $\Delta \neq 0$, the problem is no longer analytically treatable and we have to resort to numerical techniques. We use the infinite-size time-evolving block decimation (iTEBD) algorithm \cite{vidal-2007}, which implements the ideas of the density matrix renormalization group (DMRG) method \cite{schollwoeck-2005} for an infinite system. The algorithm uses an optimal matrix-product representation of the infinitely extended chain, keeping only the dominant eigenstates of the density matrix of a semi-infinite subsystem, in combination with a Suzuki-Trotter decomposition of the evolution operator.
This method is very efficient for small $t$, however, the increasing entanglement under time evolution \cite{calabrese-2007} requires to retain an exponentially growing number of eigenstates. We find that the error of our calculations behaves in a similar way to that of the finite-size DMRG algorithm and the methodology developed in Ref. \cite{gobert-2005} can be applied in order to control the accuracy \cite{barmettler-in-preparation}. By carefully estimating the {\it runaway time} via comparing results with different control parameters \cite{gobert-2005}, the absolute error in the plotted data is kept below $10^{-6}$. Using 2000 states and a second-order Suzuki-Trotter decomposition with a time step $\delta\sim 10^{-3} J^{-1}$ for large $\Delta$ and up to 7000 states with $\delta\sim 10^{-2}J^{-1}$  for small $\Delta$, an {\it intermediate} time regime $Jt\lesssim16$ can be reached, which in general far exceeds the short transient time.

An overview of the results is presented in Fig.~\ref{fig:ms_all}a. For small anisotropies we find oscillations of the order parameter similar to those in the XX-limit, but with a decay time decreasing upon approaching the isotropic point $\Delta=1$. In the easy-axis regime $\Delta>1$ of the XXZ-model, the relaxation slows down again for increasing $\Delta$, and we observe non-oscillatory behavior for $\Delta\gg1$.

\begin{figure}[t]
\includegraphics[width=\figw\textwidth]{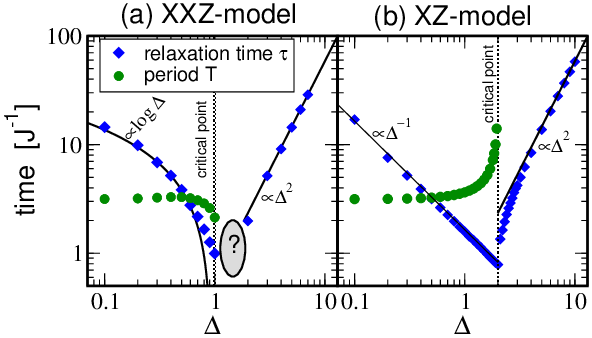}
\caption{\label{fig:relaxtime}\figpreamble Relaxation time $\tau$ and oscillation period $T=\frac{2\pi}{\omega}$ as a function of anisotropy in the XXZ- and XZ-models. Logarithmic or algebraic laws are emphasized  by solid lines. In the region close to the critical point of the XXZ-model (indicated by the question mark) it becomes impossible to extract a relaxation time from the numerical results.}
\end{figure}

Fig. \ref{fig:ms_weak} focuses on easy-plane anisotropy $0<\Delta<1$.
The results for $0<\Delta\leq0.4$ are well described, for accessible time scales, by exponentially decaying oscillations
\beq
\ms(t)\propto e^{-t/\tau}\cos(\w t +\phi)\,.
\label{eq:damping}
\eeq
The oscillation frequency is almost independent of the anisotropy, while the relaxation time $\tau$ increases with decreasing $\Delta$.  Logarithmic divergence of the relaxation time in the limit $\D\to 0$ is suggested by the fit shown in Fig. \ref{fig:relaxtime}a. The picture is less clear closer to the isotropic point. For the range $0.5\leq\Delta<1$ there appears to be an additional time scale after which the oscillations start to decay even faster than exponentially, simultaneously the period of the oscillations is reduced. Therefore, the relaxation times plotted in Fig. \ref{fig:relaxtime}a are only valid within an intermediate time window, whose width shrinks upon approaching the critical point.

For intermediate easy-axis anisotropies $1\leq\Delta\leq3$, the magnetization does not reach a stable regime within the numerically accessible time window (Fig. \ref{fig:ms_strong}a). The complicated behavior of $m_s(t)$ in this parameter range can be ascribed to the interplay of processes at all energy scales. Nevertheless, the numerical data suggest that the relaxation is fastest close to the isotropic point, in the range between $\Delta=1$ and $\Delta=1.6$. A simple generic type of behavior is recovered for large anisotropies $\Delta\gtrsim3$. The numerical data in Fig. \ref{fig:ms_strong}b indicates exponential relaxation of the staggered magnetization
\beq
\ms(t)\propto e^{-t/\tau}\,.
\label{eq:relaxation}
\eeq
The relaxation time scales roughly quadratically with $\Delta$ (Fig. \ref{fig:relaxtime}a). Oscillations do persist on top of the exponential decay, but they fade out quickly.

\paragraph{XZ-model\parend}
We now turn to the study of the XZ-Hamiltonian,
\begin{eqnarray}
H_{\text{XZ}} &=& J \sum_j \left\{ 2 \, S_j^x S_{j+1}^x + \Delta \, S_j^z S_{j+1}^z \right\}\,. \label{XZhamiltonian}
\end{eqnarray}
In this model a quantum phase transition separates two gapped phases at $\Delta_c=2$, with antiferromagnetic order in z-direction for $\Delta>\Delta_c$ and in x-direction for $\Delta<\Delta_c$. Unlike the XXZ-model, it can be easily diagonalized analytically. In order to study the staggered magnetization of the XZ-model, we have to calculate the two-spin correlation function
$
C(n,t)=(-1)^n \langle \psi_0| S^z_0(t) S^z_n(t)|\psi_0 \rangle
$
in the infinite-range limit, since $m_s^2(t)= \lim_{n\rightarrow\infty} C(n,t)$. Using standard techniques (see \cite{barouch-1} and references therein), we express this two-spin correlator as a Pfaffian, with coefficients calculated in a similar manner as for the Ising model in a transverse field \cite{sengupta-2004}.
Exploiting the light-cone effect \cite{calabrese-2007,lieb-1972}, we are able to evaluate numerically the order parameter dynamics up to times of the order of $Jt \approx 100$. The results are displayed in Fig. \ref{fig:ms_all}b.
An analytic expression can be derived for $\Delta=0$, which is given by $m_s(t)=0.5 \cos^2(J t)$.
For $\Delta<\Delta_c$, exponentially decaying oscillations 
\beq
m_s(t)\propto e^{-t/\tau}(\cos^2(\omega t)-const.)
\eeq
reproduce the numerical results at large times very well. For $\Delta \geq \Delta_c$, the staggered magnetization decays exponentially with no oscillations at large times  [Eq. (\ref{eq:relaxation})]. In contrast to the XXZ-model, the oscillation period in the XZ-model diverges at the isotropic point $\Delta=\Delta_c$ and the latter exactly marks the crossover between oscillatory and non-oscillatory behavior of $m_s(t)$.
We have extracted the relaxation times from exponential fits to the numerical data, showing a clearly pronounced minimum right at the isotropic point (see Fig. \ref{fig:relaxtime}b). The relaxation time scales as $\tau\propto\Delta^{-1}$ for $\Delta \leq \Delta_c$ and as $\tau \propto\Delta^2$ for $\Delta\gg\Delta_c$.

Apart from the numerical evaluation of the Pfaffian, we can prove rigorously that in the infinite-time limit the staggered magnetization vanishes for all anisotropies in the range $\Delta_c < \Delta < \infty$. Indeed, since the Pfaffian reduces to a Toeplitz determinant at $t\rightarrow \infty$ \cite{sengupta-2004}, we can use Szeg\"o's lemma to calculate the large-distance asymptotics of the two-spin correlator in the above-mentioned regime, obtaining for $n\gg1$, 
$\lim_{t\rightarrow\infty} C(n,t) \sim \frac{1}{4} \! \left( \frac{1+\sqrt{1-4/\Delta^2}}{2}\right)^n,$
which immediately implies that $m_s(t\rightarrow \infty)=0$.

\paragraph{Discussion\parend}

We have analyzed the dynamics of the staggered magnetization in the XXZ- and XZ-models following a quantum quench. Our main result is that in both models there is a dynamically generated relaxation rate which is fastest close to the critical point. This point also marks a crossover between oscillatory and non-oscillatory dynamics of the order parameter. The dynamics of the magnetic order parameter turns out to be a good observable for the quantitative extraction of non-trivial time scales. In general, this is not possible from other observables such as correlation functions, which reveal interesting features such as the horizon effect \cite{low-energy} but exhibit only slow relaxation dynamics \cite{spin,barmettler-in-preparation}. Furthermore, we have focused on the N\'eel state as an experimentally relevant initial condition. We point out however, that our results are generic and hold for all antiferromagnetic initial states with sufficiently small correlation length \cite{barmettler-in-preparation}.

The existence of a minimal relaxation time at the critical point is opposite to what one would expect from the phenomenon of critical slowing down of order parameter dynamics near equilibrium. In the XZ-model and the easy-axis phase of the XXZ-model, where the excitation spectrum is gapped, the effect can be understood using a phase-space argument: the relaxation of the initial state is dominated by scattering events between high-energy excitations introduced into the system through the initial state. As a result of the existence of the gap, the phase space for scattering events is restricted. This leads to an increasing relaxation time as the gap increases, whereas one expects a minimal relaxation time at the critical point, where the gap vanishes. The above argument  can not be applied directly to a quench into the gapless phase in the easy-plane regime of the XXZ-model. Rather, the situation seems to be similar to a quantum quench of the Bose-Hubbard model from a Mott insulator to a superfluid phase. In the latter case, oscillations of the superfluid order parameter have been predicted, with a damping rate that diverges at the critical point in one and two dimensions~\cite{AltmanAuerbach2002}.

The absence of a sharp signature of the quantum phase transition in the XXZ-chain prepared in a N\'eel state is in contrast with what one has in the case of the initially prepared ferromagnet with a single kink-impurity, studied for example in Ref. \cite{gobert-2005}, where the two phases are characterized by clearly distinct transport properties. We note that this initial state is much closer to the ground state of the Hamiltonian and the important energy scales are considerably smaller than in the case of the initial N\'eel state. The opening of an exponentially small gap at the phase transition is therefore more likely to be relevant.

The time evolution of an initial state which is equivalent to the N\'eel state has been recently studied by Cramer {\it et al.} \cite{cramer-2008} in the context of the one-dimensional Bose-Hubbard (BH) model with on-site repulsion $U$ as interaction parameter (the equivalence becomes apparent in the fermionic representation of the XXZ-Hamiltonian). Although the BH Hamiltonian itself and the XXZ-model share some properties in the non-interacting limit, there is one substantial difference: in the BH model at half-filling no equilibrium critical point is crossed by changing the interaction $U$ and the symmetry-broken initial state never becomes the ground state. Unlike the XXZ-model, the oscillations of the local observable in the BH model appear to be  decaying \emph{algebraically} for all values of interaction $U$ and no crossover to a non-oscillatory regime has been observed \cite{cramer-2008}. These differences point out the crucial role of the equilibrium phase transition to the reported behavior of the order parameter dynamics of the XXZ-chain.

Experimental results \cite{trotzky-2008,trotzky-private-communication} suggest that effects of density fluctuations beyond second-order magnetic exchange may be important for reproducing the dynamics in full detail. This statement is also supported by very recent numerical results \cite{barthel-2008}.  Nevertheless we expect that our main result, the existence of a minimum in the dynamically generated relaxation time close to the critical point, is insensitive to these details.

We thank D. Baeriswyl, I. Bloch, M. Lukin, M. Menteshashvili, A. M. Rey, S. Trotzky and W. Zwerger for fruitful discussions. This work was supported by SNF (P.B.), DFG - FOR 801 and the Feinberg Graduate School (M.P.), US Israel Binational Science Foundation (E.A. and E.D.) and AFOSR, CUA, DARPA, MURI, NSF DMR-0705472 (V.G. and E.D.).

\end{document}